\documentclass[aps,prb,amsmath,amssymb,reprint,superscriptaddress,preprintnumbers,showpacs,intlimits,longbibliography]{revtex4-1}
\bibpunct{[}{]}{,}{n}{}{}

\usepackage[utf8]{inputenc}

\usepackage{bm,latexsym,mathrsfs,enumerate}
\usepackage[dvipsnames]{xcolor}
\usepackage{mathtools}
\usepackage{bbm}
\usepackage[cal=boondoxo]{mathalfa}
\usepackage[breaklinks=true,unicode=true,urlcolor = blue,colorlinks = true,citecolor = blue,linkcolor = blue]{hyperref}

\usepackage{graphicx}
\graphicspath{{figs/}}
\renewcommand{\vec}[1]{\bm{#1}}
\begin{document}

\title{Magnetization induced shape transformations in flexible ferromagnetic rings}

\author{Yuri Gaididei}
\email{ybg@bitp.kiev.ua}
\affiliation{Bogolyubov Institute for Theoretical Physics of National Academy of Sciences of Ukraine, 03143 Kyiv, Ukraine}

\author{Kostiantyn V. Yershov}
\email{yershov@bitp.kiev.ua}
\affiliation{Bogolyubov Institute for Theoretical Physics of National Academy of Sciences of Ukraine, 03143 Kyiv, Ukraine}
\affiliation{National University of Kyiv-Mohyla Academy, 04655 Kyiv, Ukraine}

\author{Denis D. Sheka}
\email[Corresponding author: ]{sheka@knu.ua}
\affiliation{Taras Shevchenko National University of Kyiv, 01601 Kyiv, Ukraine}

\author{Volodymyr P. Kravchuk}
\email{vkravchuk@bitp.kiev.ua}
\affiliation{Bogolyubov Institute for Theoretical Physics of National Academy of Sciences of Ukraine, 03143 Kyiv, Ukraine}
\affiliation{Leibniz-Institut f\"ur Festk\"orper- und Werkstoffforschung, IFW Dresden, Dresden D-01171, Germany}

\author{Avadh Saxena}
\email{avadh@lanl.gov}
\affiliation{Theoretical Division and Center for Nonlinear Studies, Los Alamos National Laboratory, Los Alamos, NM 87545, USA}
\date{September 27, 2018}

\begin{abstract}
Flexible ferromagnetic rings are spin-chain magnets, in which the magnetic and mechanical subsystems are coupled. The coupling is achieved through the tangentially oriented anisotropy axis. The possibility to operate the mechanics of the nanomagnets by controlling their magnetization is an important issue for the nanorobotics applications. A minimal model for the deformable curved anisotropic Heisenberg ferromagnetic wire is proposed. An equilibrium phase diagram is constructed for the closed loop geometry: (i) A vortex state with vanishing total magnetic moment is typical for relatively large systems; in this case the wire has the form of a regular circle. (ii) A topologically trivial onion state with the planar magnetization distribution is realized in small enough systems; magnetic loop is elliptically deformed. By varying geometrical and elastic parameters a phase transition between the vortex and onion states takes place. The detailed analytical description of the phase diagram is well confirmed by numerical simulations.
\end{abstract}
\maketitle


\section{Introduction}
\label{sec:intro}

Soft magnetic materials which can change their configuration under the action of external electric or magnetic fields find applications in diverse areas of science and technology. They are used in the fabrication of materials and devices for shapeable magnetoelectronics~\cite{Makarov16,Sheng18}, programmable magnetic materials~\cite{Lum16,Kim18a} and numerous interactive human-machine interfaces \cite{Wang18,Hu18}. A possibility to control geometry of the magnet by means of the external magnetic field that acts on the magnetic subsystem opens exciting opportunities in the engineering of miniature robots~\cite{Hu18,Lum16,Kim18a}. Among different magnetically  responsive flexible materials the most studied ones are magneto-sensitive elastomers, which are composite materials of magnetic nanoparticles embedded into a nonmagnetizable polymer matrix \cite{Thevenot13,Geryak14}. Such elastomers include nanoparticle based flexible magnetic chains (wires) \cite{Singh05,Townsend14} and ribbons~\cite{Herzer13,Lopatina15}. The magnetic properties of elastomers are well described by the long range dipole-dipole interaction~\cite{Romeis17}. 

More recently, organic and molecule-based magnets have been established exhibiting different types of magnetic ordering  \cite{Ovchinnikov88,Miller88,Palacio93,Miller02,Barron08,Miller11}. Such molecule-based magnets are of great scientific interest for the development of flexible devices in the context of organic electronics and spintronics \cite{Bujak13,Miller14}. Theoretical treatment of such systems is based on the description of elastic ferromagnets, the way developed by \cite{Brown65a}. Basic models include two subsystems: the precession Landau-Lifshitz dynamics of magnetic subsystem is coupled with the Newtonian dynamics of elastic substrate \cite{Maugin99,BorovikRomanov88}. Further development of this approach for a Heisenberg magnet on elastic membranes resulted in novel effects, including periodic shrinking of the membrane due to soliton-soliton interaction \cite{Dandoloff95} and, more generally, the curvature-induced geometrical frustration in magnetic systems \cite{Saxena97,Saxena98}.

The purpose of the current study is to provide a minimal model for a \emph{flexible ferromagnet ring}, i.e. curved quasi-one-dimensional magnets with elastically deformable closed loop geometry. Here we present a detailed study of equilibrium states of the magneto-flexible ring. The geometry of the magnet was shown previously to affect the magnetic subsystem~\cite{Gaididei14,Sheka15,Gaididei17a}. Here we consider a self-consistent problem, where the inverse effect of influence of the magnetization on the magnet shape is taken into account. The coupling between the magnetic and geometrical subsystems is driven by the uniaxial anisotropy with the easy-axis oriented along the tangential direction. A small enough rigid magnetic ring is magnetized almost uniformly, forming the so-called \emph{onion state} with two domain walls \cite{Klaui03a,Guimaraes09,Sheka15}. In the case of an elastic ring the size of the domains with tangential magnetization decreases and the minimum energy is achieved by mechanical deformation of the ring shape. An opposite case of large rigid magnetic rings is known to be characterized by the flux free \emph{vortex state} \cite{Sheka15}. Since the ring shape minimizes elastic energy of any closed loop and the magnetization is everywhere tangential to the ring, the deformation of elastic ring is not favorable.  

The paper is organized as follows. In Sec.~\ref{sec:model}, we introduce a model of the magneto-flexible one-dimensional wire and discuss equations of motion for such a system. Equilibrium magnetization distributions, shape configuration of the closed loop geometry, and the phase diagram of the equilibrium states are discussed in Sec.~\ref{sec:states}. In Sec.~\ref{sec:disc}, we present final remarks and discuss the role of magnetostatic effects and excitation of zero modes. Some details concerning the analytical and numerical computations are presented in the Appendices~\ref{app:static}, \ref{app:simul}.

\section{The model}
\label{sec:model}

We consider a simple phenomenological model of a single-chain magnet of ferromagnetically coupled atoms with normalized magnetic moments $\vec{m}_i(t)$, labeled by index $i\in \overline{1,N}$. Each magnetic moment $\vec{m}_i$ is located at the point $\vec{r}_i(t)=\{x_i(t), \, y_i(t), z_i(t) \}$. We are interested in the case when the system represents a closed chain (ring), hence we impose the coordinate periodicity conditions $\vec{r}_{N+i}=\vec{r}_i$. Each magnetic moment $\vec{m}_i$ is connected with its two neighbors $\vec{m}_{i+1}$ and $\vec{m}_{i-1}$ by elastic bonds. We assume that the chain is inextensible: $|\vec{r}_{i+1}-\vec{r}_{i}|= a$ with $a$ being the lattice constant. 

In our model we take into account three contributions to the total energy of the ferromagnetic ring:

\begin{subequations} \label{eq:total_energy_discr}
\begin{equation} \label{eq:total}
\mathcal{E}=\mathcal{E}_\text{ex}+\mathcal{E}_\text{b}+\mathcal{E}_\text{an}.
\end{equation}
The first term in~\eqref{eq:total} is the Heisenberg exchange energy,
\begin{equation} \label{eq:exchange_energy}
\mathcal{E}_\text{ex} = \mathcal{J}\, \sum_{i=1}^{N}\left(\vec{m}_i-\vec{m}_{i+1}\right)^2
\end{equation}
with $\mathcal{J}>0$ being an effective exchange integral. The second term in Eq.~\eqref{eq:total} is an elastic energy~\cite{Gaididei06a,Hu13b}, which determines the change of the angle between the bond vectors $\vec{u}_{i} = \left(\vec{r}_{i+1}-\vec{r}_i\right) /a$ and $\vec{u}_{i+1} = \left(\vec{r}_{i+2} - \vec{r}_{i+1}\right) /a$,
\begin{equation} \label{eq:bending_energy}
\mathcal{E}_\text{b}=\mathcal{B}\,\sum_{i=1}^{N}\kappa_i^2,\quad \kappa^2_i\equiv \left|\vec{u}_{i+1}-\vec{u}_{i} \right|^2\!\!,
\end{equation}
where $\mathcal{B}$ is the elastic modulus of the bending rigidity (spring constant) of the chain and $\kappa_i/a$ is the curvature of the chain at the point $i$. 

The last term determines the uniaxial magnetic anisotropy contribution, 
\begin{equation} \label{eq:anisotropy_energy}
\mathcal{E}_\text{an}=-\mathcal{K}\,\sum_{i=1}^{N}\left(\vec{m}_i\cdot \vec{u}_{i}\right)^2\!,
\end{equation}
where $\mathcal{K}>0$ being the effective on-site anisotropy constant of an easy-axis type. Such kind of anisotropy is effectively induced by the dipole-dipole interaction in the thin wire~\cite{Slastikov12}.
\end{subequations}

The dynamics of the magnetic subsystem is governed by the discrete version of the Landau--Lifshitz--Gilbert equations
\begin{equation} \label{eq:LLG-discrete}
\dot{\vec{m}}_i=\frac{|\gamma|}{\mu_s} {\vec{m}}_i \times \frac{\partial \mathcal{E}}{\partial{\vec{m}}_i}+\alpha{\vec{m}}_i \times \dot{\vec{m}}_i.
\end{equation}
Here the overdot indicates a derivative with respect to time, $\alpha$ is the Gilbert magnetic damping constant, $\gamma$ is the gyromagnetic ratio, and $\mu_s$ is the magnetic moment of a magnetic site~(atom).

The dynamics of the mechanical subsystem is governed by Newton equations for atom positions $\vec{r}_i(t)$. For the sake of simplicity we will neglect inertia effects and take the equations of motion for the mechanical degrees of freedom in the form of overdamped Newton equations
\begin{subequations} \label{eq:newton}
\begin{align} \label{eq:newton_overdamped}
\frac{\partial\mathcal{R}}{\partial\dot{\vec{r}}_i} &= -\frac{\partial \mathcal{E}}{\partial\vec{r}_i},
\shortintertext{where}
\label{eq:diss_func}
\mathcal{R} &= \frac{\nu}{2} \sum_{i=1}^N {\dot{\vec{r}}_i}^2
\end{align}
\end{subequations}
is a dissipation function of the mechanical subsystem with $\nu$ being the mechanical relaxation constant. 

In what follows we limit ourselves to the case of weak anisotropy, $\mathcal{K}\ll \mathcal{J}$ and small curvature, $\kappa_i\ll1$. In this case the characteristic size of excitations $w =a \sqrt{\mathcal{J}/\mathcal{K}}$ (magnetic length) is larger than the lattice constant $a$. Thus, in the lowest approximation for the small parameter $a/w$ and weak gradients of magnetic and elastic variables, we can use the continuum approximation for the energy \eqref{eq:total_energy_discr}. The energy functional,  normalized by $\mathcal{E}_0 = \sqrt{\mathcal{JK}}$, has the following form in terms of magnetization unit vector $\vec{m}(\xi,t)$ and the tangent unit vector $\vec{u}(\xi,t)$ with $\xi=s/w$ being the normalized arc length $s$:
\begin{equation} \label{eq:total_energy}
\mathscr{E} = \int\limits_0^L\left[\vec{m}'^2 + \beta \vec{u}'^2 - \left(\vec{m}\cdot \vec{u}\right)^2\right]\mathrm{d}\xi,
\end{equation}
where $L=aN/w$ is the normalized length of the wire, $\beta=\mathcal{B}/\mathcal{J}$ is the renormalized bending parameter, and prime~($'$) denotes the derivative with respect to $\xi$. One has to note the correspondence between the exchange and bending energy terms in~\eqref{eq:total_energy}.

Spatio-temporal evolution of the system is governed by continuum equations for magnetic and elastic subsystems. Using the angular parametrization for the magnetization unit vector $\vec{m}(\xi,t)$ and the tangent unit vector $\vec{u}(\xi,t)$,
\begin{subequations} \label{eq:constrains}
\begin{align} \label{eq:m-angular}
\vec{m} &= \left(\sin\theta\cos\phi, \sin\theta\sin\phi,\cos\theta\right) ,\\
\label{eq:u-angular}
\vec{u} &= \left(\sin\psi\cos\chi, \sin\psi\sin\chi,\cos\psi\right) ,
\end{align}	
\end{subequations}
 an explicit form of the energy functional in terms of angular variables is obtained, see Appendix
\ref{app:static}.

\section{Equilibrium states of the flexible ferromagnetic ring}
\label{sec:states}

The equilibrium states of the system are determined by the minimum of the energy functional \eqref{eq:total_energy}. It corresponds to the planar magnetization distribution in a planar wire with the magnetization vector $\vec{m}(\xi)$ lying within the wire plane,
\begin{equation} \label{eq:equilibrium}
\vec{m}_0 = \left(\cos\phi_0, \sin\phi_0,0\right), \quad
\vec{u}_0 = \left(\cos\chi_0, \sin\chi_0,0\right).
\end{equation}
The corresponding azimuthal magnetic angle $\phi_0$ and the azimuthal elastic angle $\chi_0$ are determined by the set of two coupled pendulum equations~(for details, see Appendix~\ref{app:static}):
\begin{equation} \label{eq:phi_chi_stat}
2 \phi_0'' = \sin 2 \left(\phi_0-\chi_0\right),\quad 2\beta \chi_0'' =-\sin 2\left(\phi_0-\chi_0\right).
\end{equation}

\begin{figure}[t]
	\includegraphics[width=\columnwidth]{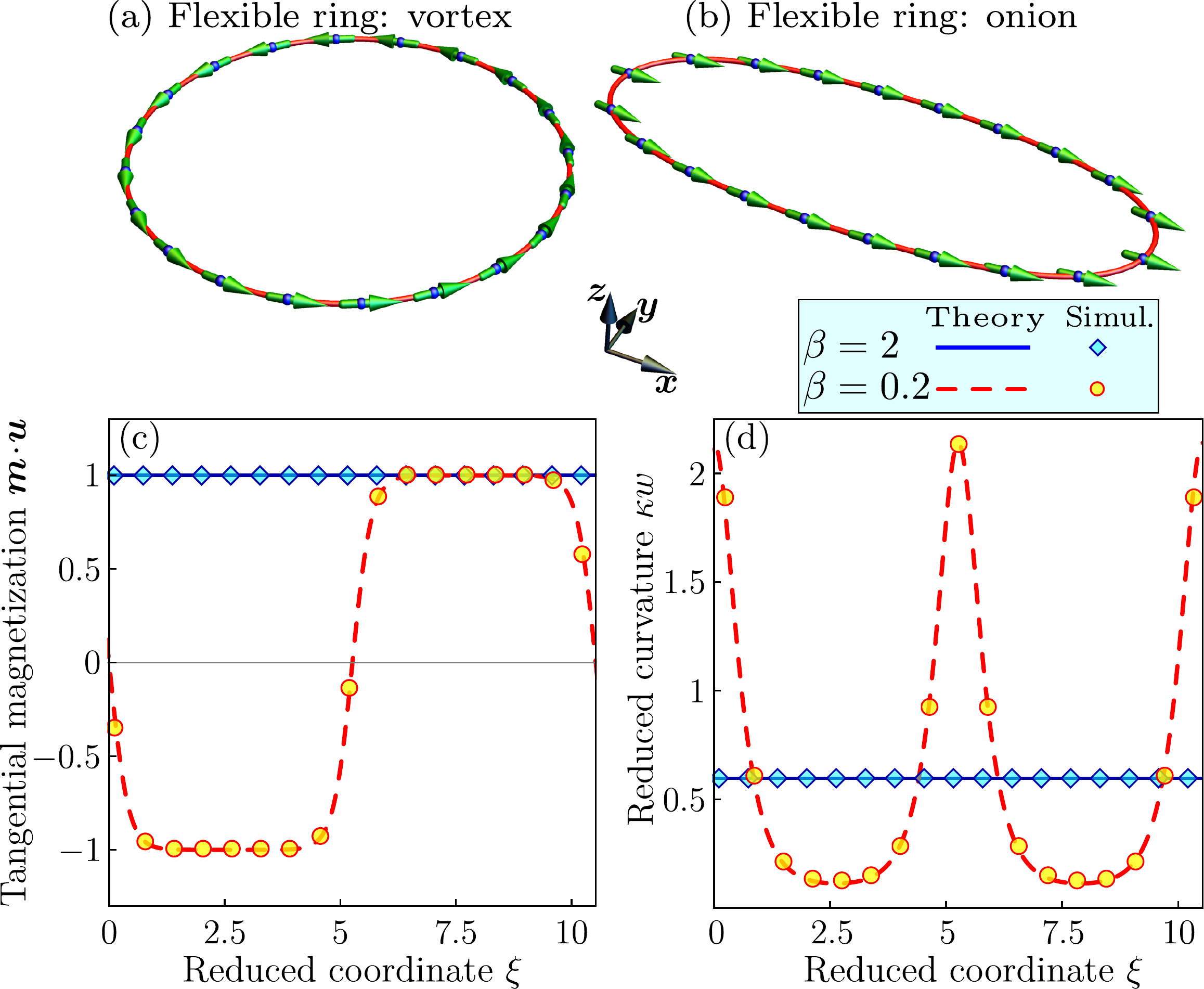}
	\caption{\label{fig:chain_states}%
		(Color online) \textbf{Equilibrium state of the flexible ring:} (a), (b) Magnetization (green arrows) and magnetic sites (blue dots) distribution obtained from numerical simulations for a chain with  $L\approx10.5$. (c), (d) Comparison of theoretical predictions~\eqref{eq:vortex_sol}, \eqref{eq:onion_sol}~(lines) and results of numerical simulations (markers). The \textit{vortex} state is obtained for $\beta=2$ and the \textit{onion} state is obtained for $\beta=0.2$.}
\end{figure}

The set of Eqs.~\eqref{eq:phi_chi_stat} with boundary conditions
\begin{equation} \label{eq:vortex_bc}
\begin{split}
\phi_0(L)=&\phi_0(0) + 2\pi, \qquad \phi'_0(L)=\phi'_0(0),\\
\chi_0(L)=&\chi_0(0) + 2\pi, \qquad \chi'_0(L)=\chi'_0(0)
\end{split}
\end{equation}
has a solution of the form
\begin{equation} \label{eq:vortex_sol}
\phi^{\text{vor}}(\xi)=\chi^{\text{vor}}(\xi)=2\pi \xi/L.
\end{equation}	 
We refer to this solution as a \emph{vortex state}~[see Fig.~\ref{fig:chain_states}(a)]: it describes the flux-free magnetization distribution in a \textit{circular} wire. It is similar to the rigid case \cite{Sheka15}. The energy of the vortex state is
\begin{equation} \label{eq:vortex_energy}
\mathscr{E}^{\text{vor}}(\beta,L) = \frac{4\pi^2}{L}\left(1+\beta\right) - L.
\end{equation}

Another type of boundary conditions
\begin{equation} \label{eq:onuion_bc}
\begin{split}
\phi_0(L)=&\phi_0(0), \qquad \phi'_0(L)=\phi'_0(0),\\
\chi_0(L)=&\chi_0(0) + 2\pi, \qquad \chi'_0(L)=\chi'_0(0)
\end{split}
\end{equation}
corresponds to the equilibrium solution
\begin{equation} \label{eq:onion_sol}
\begin{split}
\!\!\phi^{\text{on}}(\xi)	&=\frac{\beta}{1+\beta} \left[\frac{2\pi \xi}{L} - \frac{\pi}{2} - \mathrm{am}\left(\frac{4 \mathrm{K}(k)}{L}\xi, k \right)\right],\\
\!\!\chi^{\text{on}}(\xi)	&=\frac{1}{1+\beta} \left[\frac{2\pi \beta \xi}{L} + \frac{\pi}{2} + \mathrm{am} \left(\frac{4 \mathrm{K}(k)}{L}\xi, k \right)\right],
\end{split}
\end{equation}
where $\mathrm{am}(x,k)$ is the Jacobi elliptic amplitude function \cite{NIST10}. The modulus $k$ is determined by the equation
\begin{equation} \label{eq:modulus}
\sqrt{k} \mathrm{K}(k) = \frac{L}{4}\sqrt{\frac{1+\beta}{\beta}},
\end{equation}	 
where $\mathrm{K}(k)$ is the complete elliptic integral of the first kind~\cite{NIST10}.
The corresponding magnetization solution is analogous to the well-known onion state \cite{Klaui03a,Guimaraes09,Sheka15} typical for the ring geometry; hence we refer to \eqref{eq:onion_sol} as the \emph{onion state} in an \emph{elliptical} wire, see Fig.~\ref{fig:chain_states}(b). The normalized energy of the onion state reads
\begin{equation} \label{eq:onion_energy}
\!\!\! \mathscr{E}^{\mathrm{on}}(\beta,L) \!\!= \!\! \frac{4\pi^2 \beta^2}{L(1+\beta)} - \frac{L}{k} +\frac{16 \beta  \mathrm{K}(k) \mathrm{E}(k)}{L (1+\beta)} +\frac{L \mathrm{E}(k)}{ k \mathrm{K}(k)}\!,\!
\end{equation}
where $\mathrm{E}(k)$ is the complete elliptic integral of the second kind~\cite{NIST10}.

In order to verify our analytical results we performed numerical simulations, for details see~Appendix~\ref{app:simul}. The obtained numerical results confirm our analytical predictions, namely: (i) The magnetization distribution for the vortex state corresponds to the tangential direction with $\vec{m}\cdot \vec{u}=1$, which lies in the wire plane. The wire has a regular circular shape with a constant curvature $\kappa$, see Fig.~\ref{fig:chain_states}. (ii) The onion state is characterized by the planar elliptical deformation of the wire according to \eqref{eq:onion_sol} with corresponding magnetization distribution $\phi^{\text{on}}(\xi)$, see Fig.~\ref{fig:chain_states}.

\begin{figure}[t]
	\includegraphics[width=\columnwidth]{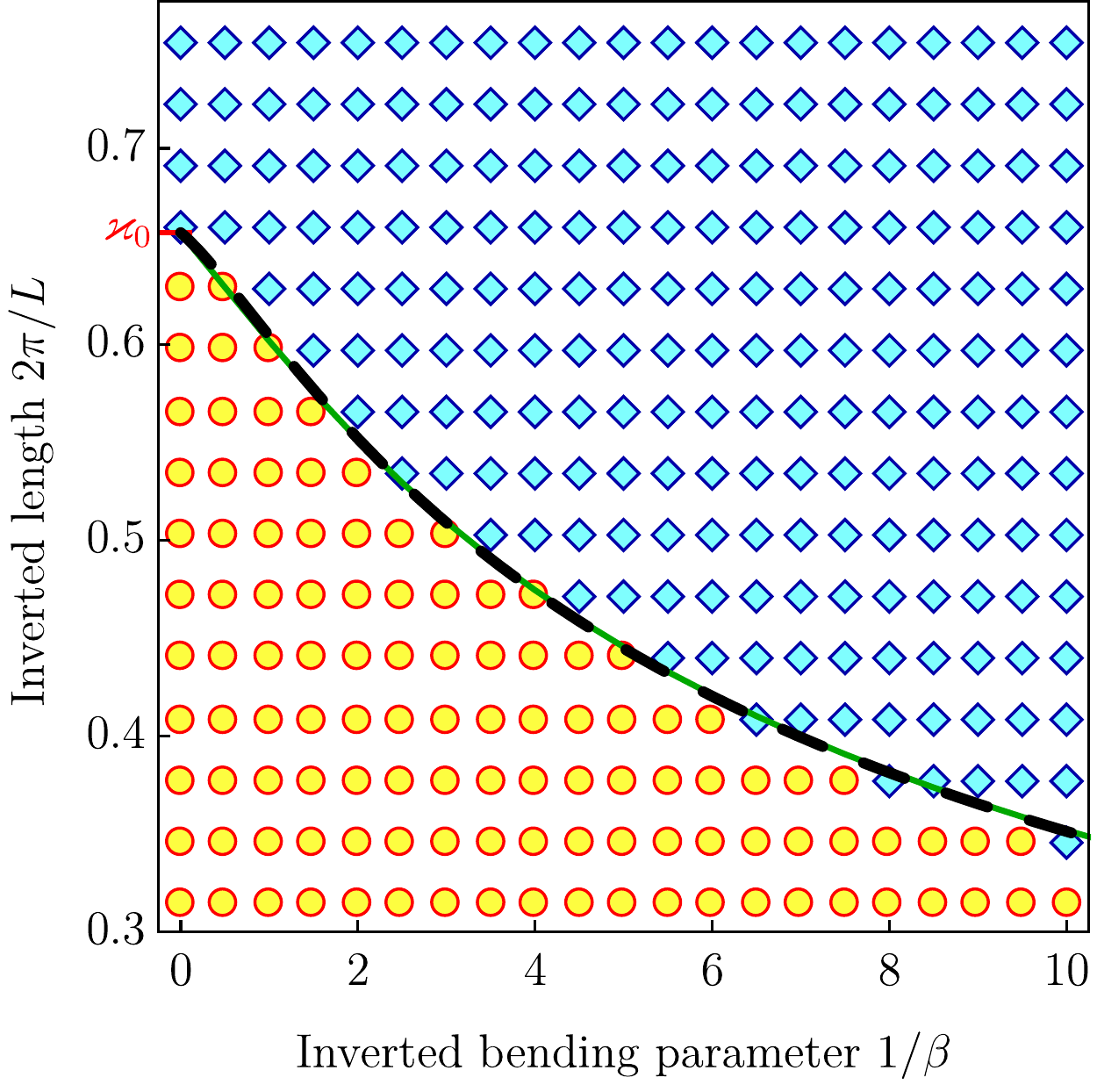}
	\caption{\label{fig:phase_diagram}
		(Color online) \textbf{Phase diagram of the equilibrium magnetization and shape states for the flexible ferromagnet.} Symbols correspond to simulation: yellow circles correspond to the vortex magnetization distribution with circular shape of the wire; blue diamonds correspond to the onion magnetization distribution with \emph{elliptical} wire shape. Green solid line describes the boundary $L_b(\beta)$ between the vortex and the onion states plotted with the prediction~\eqref{eq:states_boundary} and the dashed line is its fitting by~\eqref{eq:states_boundary_trial}.}
\end{figure}

Next, we summarize the results of the equilibrium states of the system: both the magnetization distribution and the wire configuration. By comparing energies of different states, we find the energetically preferable states for different values of the normalized wire length $L$ and the bending elasticity parameter $\beta$. The resulting phase diagram is presented in Fig.~\ref{fig:phase_diagram}. There are two phases: (i) The vortex state is realized for relatively large $L$, when $L> L_b(\beta)$. In such a state, the magnetization is directed in the tangential direction to the wire of a circular shape, in accordance with \eqref{eq:vortex_sol}, see Fig.~\ref{fig:chain_states}(a). (ii) The onion state is energetically preferable, when $L < L_b(\beta)$. The magnetization distribution is inhomogeneous and the wire has elliptical shape in accordance with~\eqref{eq:onion_sol}, see Fig.~\ref{fig:chain_states}(b).

The boundary between the two phases $L_b = L_b(\beta)$ can be derived by using the condition $\mathscr{E}^{\text{vor}}\left(\beta,L_b \right) = \mathscr{E}^{\text{on}}\left(\beta,L_b \right)$:
\begin{subequations}\label{eq:states_boundary}
	\begin{align}
		\frac{2\pi}{L_b}=\frac{\pi}{2}\frac{\sqrt{1+1/\beta}}{\sqrt{k_0}\mathrm{K}(k_0)},
	\end{align}
where $k_0$ is the solution of equation
	\begin{align}
		2\mathrm{K}(k_0)\mathrm{E}(k_0)+\mathrm{K}^2(k_0)(k_0-1)=\frac{\pi^2}{4}\left(2+1/\beta\right).
	\end{align}
\end{subequations}
In the case of a rigid ring ($\beta\to\infty$) one gets a limit value ${2\pi}/{L_b}=\varkappa_0\approx0.657$, which corresponds to the reduced curvature of a circular-shaped wire; such a critical parameter is known to separate two different equilibrium magnetization states of the ring~\cite{Sheka15}. An opposite limit case of extremely flexible ring results in the boundary curve $2\pi/L_b \approx \left(4/\pi\right)\sqrt{\beta}$.
For the approximate description of the boundary dependence~\eqref{eq:states_boundary} we can use the fitting function
\begin{equation} \label{eq:states_boundary_trial}
\frac{2\pi}{L_b^\star}=\varkappa_0 \frac{\sqrt{c^2 +1/\beta}}{c+\pi \varkappa_0/\left(4\beta\right)},\quad c\approx 1.106.
\end{equation}
Providing the asymptotically correct behavior, this approximation reproduces results \eqref{eq:states_boundary} with an accuracy of about $6\times 10^{-4}$.

\section{Discussion}
\label{sec:disc}

We have performed a detailed study of the statics of ferromagnetic rings in the context of soft condensed matter. Specifically, we proposed a minimal model \eqref{eq:total_energy} for the curved anisotropic Heisenberg ferromagnet on an elastically deformable curved wire.

\begin{figure*}[t]
	\includegraphics[width=\textwidth]{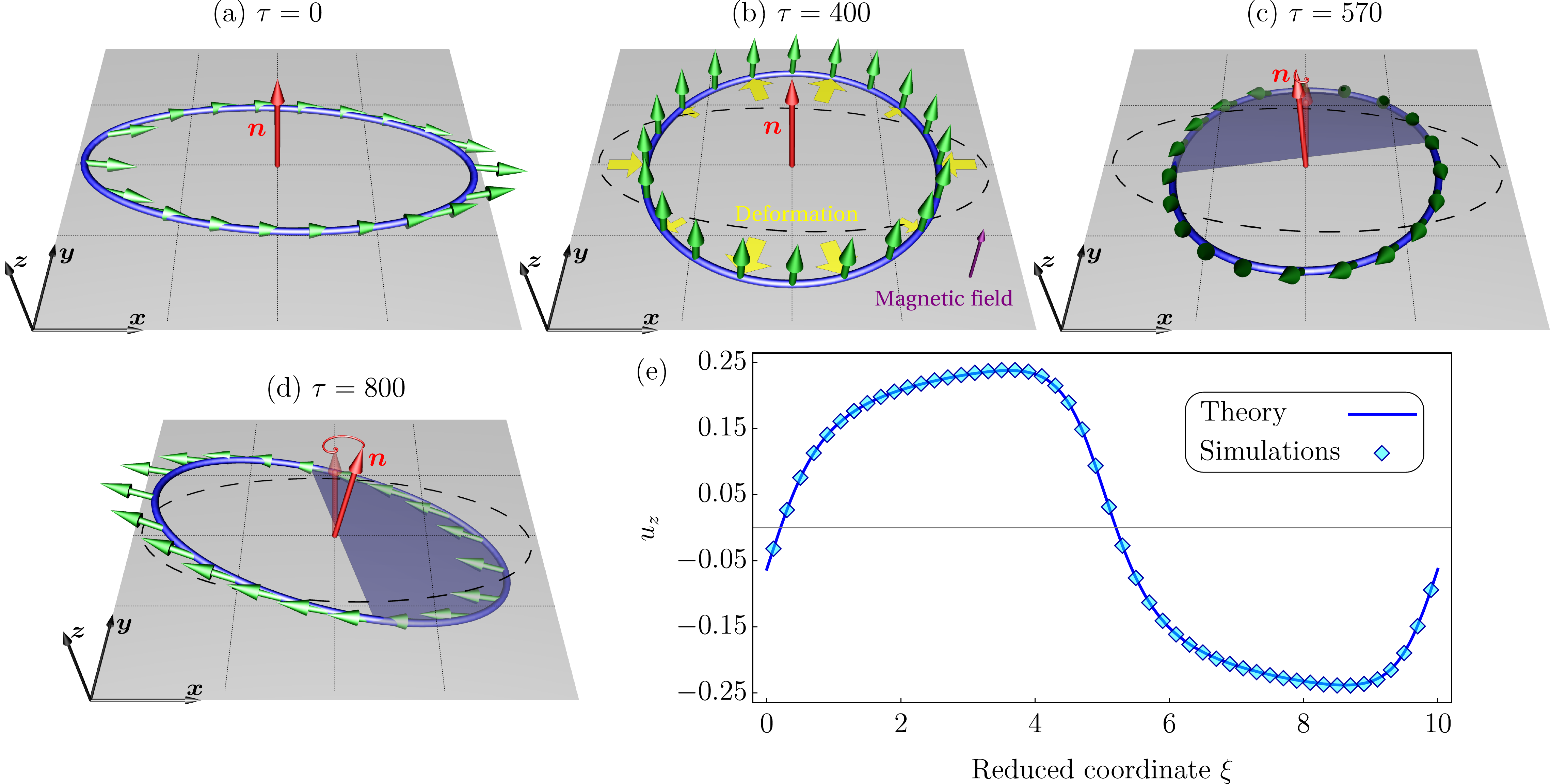}
	\caption{\label{fig:inclination}
		(Color online) \textbf{Rotation of the loop plane:} (a)-(d) Magnetic and elastic configuration of the flexible system according to numerical simulations with $L\approx 10$, $\beta=0.5$, $\alpha=0.01$, and $\nu=0.01\mu_s/\left(a^2|\gamma|\right)$: (a) initial equilibrium configuration; (b) deformed system under the action of external magnetic field and mechanical stress with temporal profile $f(\tau)$, described by Eq.~\eqref{eq:f(t)} with $\tau_1=50$, $\tau_2=550$, and $\delta=5$; (c) intermediate state of the ring during the rotation; (d) relaxed configuration after switching off external influence.  Red arrow determines the direction of the normal vector $\vec{n}$ to the ring plane (in Fig.~(d) the normal vector $\vec{n}$ does not lie in the $x0z$ plane). Black dashed ellipse-like ring in Figs.~(b)-(d) corresponds to the initial shape of the ring (a). Purple and yellow arrows in Fig.~(b) determine the direction of the external magnetic field~$f(\tau)\vec{b}$ and deformation $f(\tau)\left[\vec{r}_i(\tau)-\vec{r}_i(0)\right]$, respectively. (e) The comparison of theoretical predictions \eqref{eq:uz-modes} and the results of numerical simulations.  The corresponding dynamics is illustrated in the supplementary video~\cite{Note1}.
	}
\end{figure*}

First we discuss the symmetry of the model and its consequences. The energy functional \eqref{eq:total_energy} is invariant with respect to the joint rotation of elastic $\vec{u}$-vector and magnetization $\vec{m}$-vector through an angle $\gamma$ about some axis given by $\vec{g} = \left(\sin \psi_g\cos \chi_g, \sin \psi_g \sin\chi_g,\cos\psi_g\right)$,
\begin{equation}\label{eq:symmetry}
\vec{u}(\xi) \rightarrow U^{\vec{g}}(\gamma)\vec{u}(\xi),\quad \vec{m}(\xi) \rightarrow U^{\vec{g}}(\gamma)\vec{m}(\xi).
\end{equation}
For an explicit form of the rotation matrix $U^{\vec{g}}(\gamma)$, see Appendix \ref{app:linear}. The consequence of this symmetry is the appearance of normal modes with zero frequency (zero modes) on the background of equilibrium states \eqref{eq:equilibrium}. The corresponding elastic zero mode, as a linear excitation $\Delta \vec{u}$ on the background of the static solution $\vec{u}_0$, i.e. $\Delta \vec{u} = \vec{u}-\vec{u}_0$, is determined by the infinitesimal rotation:
\begin{equation} \label{eq:Delta-u}
\Delta \vec{u}\propto \partial_\gamma U^{\vec{g}}(\gamma)\vec{u}_0(\xi)\Biggr |_{\gamma=0} \!\!\! = 
\begin{Vmatrix}
-\cos\psi_g \sin\chi_0\\
\cos\psi_g \cos\chi_0\\
\sin\psi_g \sin\left(\chi_0-\chi_g\right)\\
\end{Vmatrix}.
\end{equation}
In the same way one can derive the magnetic zero mode, $\Delta \vec{m}\propto \partial_\gamma U^{\vec{g}}(\gamma)\vec{m}_0(\xi)\Bigr |_{\gamma=0}$.

In terms of angular variables, zero mode solutions read
\begin{equation} \label{eq:zero_modes}
\begin{aligned}
\Delta \psi 	&= c_1\sin\left(\chi_0-\chi_g\right), & \Delta \chi &= c_2,\\
\Delta \theta 	&= c_1\sin\left(\chi_0-\chi_g\right), & \Delta \phi &= c_2,
\end{aligned}
\end{equation} 
where $\chi_g$ is a constant angle; $c_1$ and $c_2$ describe amplitudes of the corresponding modes, see Appendix \ref{app:linear} for the details.

By exciting zero modes one can swing the loop plane through some angle. We have checked this idea by means of numerical simulations. Specifically, we choose the system parameters which correspond to the onion state, for details see Appendix \ref{app:zero_modes}. We relax the system to its equilibrium magnetization state with elliptical shape of the elastic subsystem. By applying an external mechanical force together with a magnetic field, we deform both the magnetic and the elastic structure of the flexible magnetic ring. After switching off the external influence, we observe numerically that the system relaxes to the equilibrium state, which is accompanied by a loop plane swing, see Fig.~\ref{fig:inclination} and supplementary movie~\footnote{See Supplemental Movie at \url{https://youtu.be/f-7y0wdN98E}}.

In order to estimate the effect we compare the numerically obtained dependence $u_z(\xi)=z'/w$ with the linear mode profile for the out-of-plane component of the elastic unit vector:
\begin{equation} \label{eq:uz-modes}
u_z(\xi) = \cos\psi = u_0 \sin \left(\chi^{\text{on}}(\xi) - \chi_g\right),
\end{equation}
where $\chi_g$ is the azimuthal angle of the vector $\vec{g}$. One can see from Fig.~\ref{fig:inclination}, that the zero-mode solution \eqref{eq:uz-modes} fits to a good accuracy with the simulation data, see Appendix~\ref{app:zero_modes} for details.

Next, let us discuss how our model can be generalized taking into account the long-range magnetostatic effects. The nonlocal magnetostatic interaction for thin wires of circular and square cross sections is known~\cite{Slastikov12} to be reduced to a local effective easy-tangential anisotropy $\mathcal{K}^\text{ms}=\pi \mu_s^2/a^3$. It is important that such a conclusion survives for the case of curved wires~\cite{Slastikov12}. Thus the magnetostatic interaction can be taken into account as an additional anisotropy. In this case, the magnetostatic effects can be taken into account by a simple redefinition of the anisotropy constants, leading to a new magnetic length,
\begin{equation}\label{eq:effective_length}
	\begin{split}
		&\mathcal{K}\to \mathcal{K}^\text{eff}=\mathcal{K}+\mathcal{K}^\text{ms},\\
		&w\to w^\text{eff} = a\sqrt{\frac{\mathcal{J} }{\mathcal{K}^\text{eff}}}=\frac{2\ell}{\sqrt{1+2\mathcal{Q}}},
	\end{split}
\end{equation}
where $\ell=a\sqrt{\mathcal{J}/\left(4\mathcal{K}^\text{ms}\right)}$ is an exchange length and $\mathcal{Q}=\mathcal{K}/\left(2\mathcal{K}^\text{ms}\right)$ is a quality factor~\cite{Hubert98}.

In order to check our predictions about the effective anisotropy, we performed numerical simulations taking into account the nonlocal dipolar interaction~(for details, see Appendix~\ref{app:simul}). We performed simulations for flexible and magnetically \textit{soft}~($\mathcal{Q}=0$) wires including the dipolar interaction. According to~\eqref{eq:effective_length}, we get $w^\text{eff}=2\ell$. Simulation data are presented in Fig.~\ref{fig:ms} for the renormalized length of the ring, according to
\begin{equation} \label{eq:effective_curvature}
L\to L^\text{eff}=\frac{aN}{w^\text{eff}}.
\end{equation}

Note that the local approach is known \cite{Kravchuk07} to provide an under-estimate of the magnetostatic energy of the onion state, while for the vortex state it gives correct results. The effective anisotropy approach~\eqref{eq:effective_length} does not take into account the long range part of the magnetostatic interaction. The consequence of this effect is the shifting down of the analytically estimated boundary between vortex and onion states \eqref{eq:states_boundary} in comparison with simulations, see Fig.~\ref{fig:ms}(c).

In conclusion, we have presented a minimal model for studying a flexible magnetic wire. The ground states of the system essentially depend on geometric, magnetic, and elastic parameters. Depending on the parameters, one can distinguish two different states: the onion state with the quasi-uniform magnetization typical for small enough rings; the vortex state with the magnetization oriented tangential to the wire is preferable in the opposite case. We have calculated the phase diagram of possible states in a wide range of bending constants $\beta$ and normalized ring lengths $L$.

\begin{acknowledgements}
The authors are grateful to Dr. Oleksandr V. Pylypovskyi and Dr. Oleksii M. Volkov for fruitful discussions. D.~D.~S. thanks Helmholtz-Zentrum Dresden-Rossendorf e.~V., where part of this work was performed, for their kind hospitality and acknowledges the support from the Alexander von Humboldt Foundation (Research Group Linkage Programme). In part, this work was supported by the Program of Fundamental Research of the Department of Physics and Astronomy of the National Academy of Sciences of Ukraine (Project No.~0116U003192) and the U.S. Department of Energy. 

\end{acknowledgements}


\appendix


\section{Static solutions}
\label{app:static}

In order to analyze static configurations, let us start with the total energy of the system \eqref{eq:total_energy}. Using angular parametrization \eqref{eq:constrains} one gets the energy functional in the following form,
\begin{widetext}
\begin{equation} \label{eq:total_energy_angular}
E = \int\limits_0^L\left\{\theta'^2+\sin^2\theta\,\phi'^2
+ \beta \left(\psi'^2 + \sin^2\psi\, \chi'^2\right) -  \left[\sin\theta\sin\psi\cos(\phi-\chi) + \cos\theta \cos\psi \right]^2\right\}\mathrm{d}\xi.
\end{equation}

Minimization of the energy functional \eqref{eq:total_energy_angular} results in static equations:
\begin{subequations} \label{eq:continuum-statics}
\begin{align} \label{eq:continuum-statics-theta}
&\theta''-\sin\theta\cos\theta \phi'^2 + \left[\sin\theta\,\sin\psi\cos(\phi-\chi) + \cos\theta \cos\psi\right]\, \left[\cos\theta\,\sin\psi \cos(\phi-\chi)-\sin\theta \cos\psi\right]=0,\\
\label{eq:continuum-statics-phi} %
&\beta \left[\psi''-\sin\psi \cos\psi \chi'^2\right] + \left[\sin\theta\,\sin\psi\cos(\phi-\chi) + \cos\theta \cos\psi\right]\, \left[\sin\theta\,\cos\psi \cos(\phi-\chi) - \cos\theta \sin\psi\right]=0,\\
\label{eq:continuum-statics-Psi} %
& \left(\sin^2\theta \phi'\right)' -\left[\sin\theta\,\sin\psi\cos(\phi-\chi) + \cos\theta \cos\psi\right]\,\sin\theta\,\sin\psi\sin(\phi-\chi)=0,\\
\label{eq:continuum-statics-chi} %
&
\beta \left(\sin^2\psi \chi'\right)' + \left[\sin\theta\,\sin\psi\cos(\phi-\chi) + \cos\theta \cos\psi\right]\,\sin\theta\,\sin\psi\sin(\phi-\chi)=0.
\end{align}
\end{subequations}
\end{widetext}
The set of static equations \eqref{eq:continuum-statics} has a solution in the form of the planar wire (elastic polar angle $\psi=\pi/2$) with the planar magnetization distribution (magnetization polar angle $\theta=\pi/2$). This results in a set of two equations for azimuthal magnetic angle $\phi(\xi)$ and azimuthal elastic angle $\chi(\xi)$, see Eqs.~\eqref{eq:phi_chi_stat}.


\section{Numerical simulations}
\label{app:simul}

In order to verify our analytical results we numerically simulate the magnetization dynamics of a flexible chain of discrete magnetic moments $\vec{m}_i(t)$ located in the positions $\vec{r}_i(t)$ with $i\in\overline{1,N}$. For the position $\vec{r}_i$ the periodic boundary condition was applied, $\vec{r}_1=\vec{r}_{N+1}$.

The dynamics of the magnetic subsystem is described by the Landau--Lifshitz equations
\begin{subequations} \label{eq:LLG_Newt_eq}
	\begin{equation}\label{eq:LLG_mod}
	\frac{\mathrm{d}\vec{m}_i}{\mathrm{d}\tau}
	=\vec{m}_i \times \frac{\partial\mathcal{H}}{\partial\vec{m}_i}+\alpha\, \vec{m}_i \times \left[\vec{m}_i \times \frac{\partial\mathcal{H}}{\partial\vec{m}_i}\right],
	\end{equation}
	while the dynamics of the mechanical subsystem is described by the over-damped Newton equations
	\begin{equation}\label{eq:Newt_mod}
	\eta\frac{\mathrm{d}\vec{r}_i}{\mathrm{d}\tau}=-\frac{\partial\mathcal{H}}{\partial\vec{r}_i},
	\end{equation}
\end{subequations}
where $\tau=\omega_0 t$ is a reduced time with $\omega_0=4\pi|\gamma| \mu_s/a^3$, $\alpha$ and $\eta=\nu|\gamma|/\mu_s$ are damping coefficients, $\mathcal{H}$ is the dimensionless energy normalized by $4\pi \mu_s^2/a^3$. We consider five contributions to the energy of the system:
\begin{subequations} \label{eq:tot-energy-eq-mod}
	\begin{equation} \label{eq:tot-energy-mod}
	\mathcal{H} = \mathcal{H}_\text{ex} + \mathcal{H}_\text{an} + \mathcal{H}_\text{d} + \mathcal{H}_\text{b} + \mathcal{H}_\text{str}.
	\end{equation}
	The first term in Eq.~\eqref{eq:tot-energy-mod} is the exchange energy
	\begin{equation} \label{eq:energy-exch-mod}
	\mathcal{H}_{\text{ex}} = -2\frac{\ell^2}{a^2} \sum\limits_{i=1}^{N-1} \vec{m}_i \cdot \vec{m}_{i+1}.
	\end{equation}
	The second term determines the uniaxial anisotropy contribution
	\begin{equation} \label{eq:energy-anis-mod}
	\mathcal{H}_{\text{an}} =-\frac{\mathcal{Q}}{2}\sum\limits_{i=1}^{N} (\vec{m}_i \cdot \vec{u}_i)^2.
	\end{equation}
	The third term determines the dipolar interaction
	\begin{equation} \label{eq:energy-dip-mod}
	\mathcal{H}_{\text{d}} =\frac{a^3}{8\pi}\sideset{}{'}\sum\limits_{i,j} \left[\frac{\vec{m}_i\cdot\vec{m}_j}{|\vec{r}_{ij}|^3}-3\frac{\left(\vec{m}_i\cdot\vec{r}_{ij}\right)\left(\vec{m}_j\cdot\vec{r}_{ij}\right)}{|\vec{r}_{ij}|^5}\right],
	\end{equation}
	where $\vec{r}_{ij}=\vec{r}_i-\vec{r}_j$. The fourth term determines the bending potential
	\begin{equation} \label{eq:energy-bend-mod}
	\mathcal{H}_{\text{b}}=\beta\frac{\ell^2}{a^2} \sum\limits_{i=1}^{N}\left|\vec{u}_{i+1}-\vec{u}_{i}\right|^2.
	\end{equation}
	The last term in~\eqref{eq:tot-energy-mod} determines the stretching energy
	\begin{equation} \label{eq:energy-str-mod}
	\mathcal{H}_{\text{str}}=\frac{\Lambda}{a^2}  \sum\limits_{i=1}^{N}\left(\left|\vec{r}_{i}-\vec{r}_{i+1}\right|-a\right)^2,
	\end{equation}
	where $\Lambda\gg 1$ determines the stretching rigidity constant.
\end{subequations}

The dynamical problem is considered as a set of $6N$ ordinary differential equations \eqref{eq:LLG_Newt_eq} with respect to $6N$ unknown functions $m_i^{\text{x}}(t),\,m_i^{\text{y}}(t),\,m_i^{\text{z}}(t)$, $r_i^{\text{x}}(t),\,r_i^{\text{y}}(t),\,r_i^{\text{z}}(t)$. For given initial conditions the set~\eqref{eq:LLG_Newt_eq} is integrated numerically. During the integration process condition $|\vec{m}_i(t)|=1$ is controlled.  

\subsection{Equilibrium states of a flexible chain}
\label{app:ES_ring}

\begin{figure*}[t]
	\includegraphics[width=\textwidth]{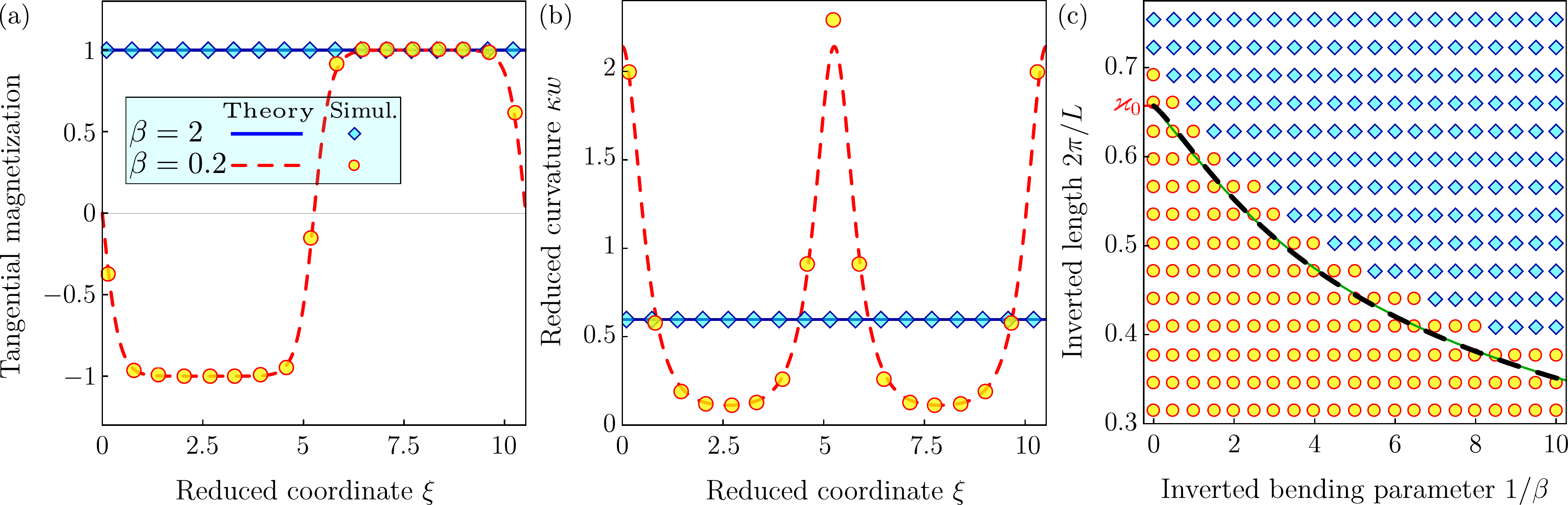}
	\caption{\label{fig:ms}
		(Color online) \textbf{Flexible ferromagnetic rings with inclusion of the dipole-dipole interaction:} (a), (b) Comparison of theoretical predictions~\eqref{eq:vortex_sol}, \eqref{eq:onion_sol}~(lines) and results of numerical simulations (markers). The \textit{vortex} state is obtained for $\beta=2$ and the \textit{onion} state is obtained for $\beta=0.2$. (c) Phase diagram of the equilibrium magnetization and shape states for the flexible ferromagnet. Symbols correspond to simulation: yellow circles correspond to the vortex magnetization distribution with circular shape of the wire; blue diamonds to the onion magnetization distribution with \emph{elliptical} wire shape. Green solid line describes the boundary $L_b(\beta)$ between the vortex and the onion states plotted with the prediction~\eqref{eq:states_boundary} and the dashed line is its fitting by~\eqref{eq:states_boundary_trial}.
	}
\end{figure*}

We considered the ring-shaped chain with  length $L=100a$ and the quality factor $\mathcal{Q}=2$. The exchange length was varied in the range $w\in\left[5a,12a\right]$ with the step $\Delta w=0.5a$. The ring-shaped form in $xy$ plane was fixed as an initial chain-units distribution for all cases.

In order to find the equilibrium state of a flexible ring we performed the integration of \eqref{eq:LLG_Newt_eq} with damping coefficients $\alpha=0.01$ and $\nu=0.01\mu_s/\left(a^2|\gamma|\right)$ on a long time interval.  The numerical experiment consists of two steps. First, we simulate the anisotropic flexible ferromagnetic chain for five different initial magnetization distributions, namely the vortex, onion, normal, and two random states.  The final static state with the lowest energy is considered to be the equilibrium state. We obtain that in a ring-shaped chain we can realize only two magnetization distributions, namely vortex and onion. Additionally, we found that the onion state deforms the circular shape of the chain into the elliptical shape, while the vortex state preserves the circular shape of the chain, see Fig.~\ref{fig:chain_states}.

At the second step, we performed numerical simulation of the isotropic~($\mathcal{Q}=0$) flexible ferromagnetic chain including the dipolar interaction. In this case, the dipolar effects can be taken into account by a simple redefinition of the anisotropy constants~\cite{Slastikov12}, leading to a new magnetic length~\eqref{eq:effective_length}.

Simulations with dipolar interactions were performed for the exchange length $w^\text{eff}\in\left[5a,12a\right]$. Equilibrium states are studied in the same way as for the case of anisotropic chains. Resulting phase diagrams of equilibrium states are plotted in Fig.~\ref{fig:phase_diagram}.

\subsection{Zero modes}
\label{app:zero_modes}

Our theoretical treatment predicts the effect of the loop plane swing by exciting zero modes. In order to realize this effect we performed the following numerical simulations: Initially we relaxed the system to the ground onion state, see Fig.~\ref{fig:inclination}(a). By applying an external force and magnetic field pulses we deformed both the elastic subsystem and the magnetic texture, see  Fig.~\ref{fig:inclination}(b). After switching off the external excitation the system relaxes to the equilibrium state, which is accompanied by the swing of the loop plane, see Fig.~\ref{fig:inclination}(c),(d). To be more specific, the external elastic deformation is modeled by the potential
\begin{equation} \label{eq:circle_pot}
\begin{aligned}
\mathcal{H}_\textsc{c} &= f(\tau)\frac{\rho}{a^2}\sum_{i=1}^{N}|\vec{r}_i-\vec{r}_i^\textsc{c}|^2,\\ 
\vec{r}_i^\textsc{c} &= \frac{aN}{2\pi}\left(\cos\frac{2\pi i}{N},\sin\frac{2\pi i}{N},0\right),
\end{aligned}
\end{equation}
where $\rho>0$. Here the function
\begin{equation} \label{eq:f(t)}
f(\tau)=\frac12\left(\tanh\frac{\tau-\tau_1}{\delta}-\tanh\frac{\tau-\tau_2}{\delta}\right)
\end{equation}
determines the temporal profile of the external potential. Here $\tau_1$, $\tau_2$, and $\delta$ are pulse parameters. This potential deforms the configuration of the chain from elliptical to the circular one. Potential~\eqref{eq:circle_pot} results in the force which acts on the $i$-th node~(atom)
\begin{subequations}\label{eq:ring_force}
\begin{equation}\label{eq:force}
\vec{\mathcal{F}}_i=\vec{\mathcal{F}}_i^\textsc{c}+\vec{\mathcal{F}}_i^\text{def}.
\end{equation}
The first term in Eq.~\eqref{eq:force} is a reaction force
\begin{equation}\label{eq:force_c}
\vec{\mathcal{F}}_i^\textsc{c}=2f(\tau)\frac{\rho}{a^2}\left[\vec{r}_i^\textsc{c}-\vec{r}_i(0)\right].
\end{equation}
The second term determines a deformation force contribution
\begin{equation}\label{eq:force_def}
\vec{\mathcal{F}}_i^\text{def}=-2f(\tau)\frac{\rho}{a^2}\left[\vec{r}_i(\tau)-\vec{r}_i(0)\right].
\end{equation}
\end{subequations}

The interaction of magnetic subsystem with external magnetic field was taken into account as
\begin{equation}\label{eq:zeeman_pot}
\mathcal{H}_\text{z}=-f(\tau)\sum_{i=1}^{N}\left(\vec{m}_i\cdot\vec{b}\right),\quad \vec{b}=b\left(\cos\vartheta,0,\sin\vartheta\right),
\end{equation}
where $b=Ba^3/\left(4\pi\mu_s\right)$ is the amplitude of the magnetic field and $\vartheta=17\pi/36$. The magnetic field is applied at the angle $\vartheta$ to the $xy$ plane in order to avoid the transition to the metastable states. Yellow and purple arrows in Fig.~\ref{fig:inclination}(b) determine the direction of deformation force~$\vec{\mathcal{F}}_i^\text{def}$ and magnetic field~$f(\tau)\vec{b}$, respectively.

The magnetic ring with onion magnetization distribution and elliptical shape configuration after this simulation turned out to be the original $xy$ plane. Figure~\ref{fig:inclination} demonstrates the final dependence $u_z(\xi)$ which the chain units acquire as a result of rotation. The results of numerical simulation are fitted well by Eq.~\eqref{eq:uz-modes} within an accuracy of about $5\times 10^{-3}$; the fitting parameters being $u_0\approx0.24$ and $\chi_g\approx \sqrt{3}$.

It is important to mention here that the system with vortex magnetization distribution and circular configuration stayed in the $xy$ plane.


\section{Rotation of the elastic unit vector $u$}
\label{app:linear}

Let us consider the rotation of some vector through an angle $\gamma$ about the axis
\begin{equation} \label{eq:g}
\vec{g} = \left(
\begin{matrix}
g_x\\ g_y\\ g_z
\end{matrix}\right) =\left(
\begin{matrix}
\sin \psi_g\cos \chi_g \\ \sin \psi_g \sin\chi_g \\ \cos\psi_g
\end{matrix}\right).
\end{equation}
The rotation matrix can be written as~\cite{Koks06}
\begin{subequations} \label{eq:Uij}
	\begin{equation} \label{eq:U}
	U^{\vec{g}}(\gamma)=\left(1-\cos\gamma\right)\vec{g}\vec{g}^\textsc{t}+\cos\gamma\, \mathbbm{1}+\sin\gamma\, g^\times,
	\end{equation}
	where $\mathbbm{1}$ is the $3 \times 3$ identity matrix and
	\begin{equation}\label{eq:g_matrix}
	g^\times=\left(\begin{matrix}
	0 & -g_z & g_y\\
	g_z & 0 & -g_x\\
	-g_y & g_x & 0 
	\end{matrix}\right).
	\end{equation}
\end{subequations}
We consider small deviations of the elastic unit vector from the static solution, $\Delta \vec{u}= \vec{u}-\vec{u}_0$. In the same way we introduce small deviations of the angular variables, $\Delta \psi = \psi-\pi/2$ and $\Delta \chi= \chi-\chi_0$. By linearizing in $\Delta\psi$ and $\Delta\chi$, we can represent $\Delta \vec{u} = \left( -\Delta \chi \sin\chi_0,\ \Delta \chi \cos\chi_0,\ -\Delta \psi \right)$. Now, by comparing the above expression for $\Delta\vec{u}$ with \eqref{eq:Delta-u}, we get the amplitude for the zero eigenmode in the form \eqref{eq:zero_modes}.



%

\end{document}